%% file: proceedings.tex
\documentclass{sigchi}

\usepackage{balance}  
\usepackage[T1]{fontenc}
\usepackage{txfonts}
\usepackage{mathptmx}
\usepackage[pdftex]{hyperref}
\usepackage{color}
\usepackage{tcolorbox}
\usepackage{booktabs}
\usepackage{textcomp}
\usepackage{enumitem}
\usepackage{graphicx}
\usepackage{caption}
\usepackage{subcaption}
\usepackage{diagbox}
\usepackage{hyperref}

\usepackage{microtype} 
\usepackage{ccicons}  

\usepackage{todonotes}

\usepackage{multicol}
\usepackage{cuted}
\usepackage[font=small,labelfont=bf,tableposition=top]{caption}

\emergencystretch=\maxdimen
\toappear{\large{Demo video:\\ \href{https://youtu.be/HlFFL0eH3sc}{https://youtu.be/HlFFL0eH3sc} } }

\def\plaintitle{Enhancing Email Functionality using Late Bound Content}

\def\emptyauthor{}
\def\plainkeywords{Authors' choice; of terms; separated; by
  semicolons; include commas, within terms only; required.}

\makeatletter
\def\url@leostyle{%
  \@ifundefined{selectfont}{
    \def\UrlFont{\sf}
  }{
    \def\UrlFont{\small\bf\ttfamily}
  }}
\makeatother
\def\pprw{8.5in}
\def\pprh{11in}

\definecolor{linkColor}{RGB}{6,125,233}
\hypersetup{%
  pdftitle={\plaintitle},
  pdfauthor={\emptyauthor},
  pdfkeywords={\plainkeywords},
  bookmarksnumbered,
  pdfstartview={FitH},
  colorlinks,
  citecolor=black,
  filecolor=black,
  linkcolor=black,
  urlcolor=linkColor,
  breaklinks=true,
}

\usepackage{color}
\definecolor{orange}{RGB}{255,127,0}
\definecolor{limegreen}{RGB}{50, 205, 50}
\definecolor{violet}{RGB}{148,0,211}

\usepackage[utf8]{inputenc}

\newif\ifCOMMENTS
\COMMENTStrue

\ifCOMMENTS

\else

\fi

\begin{document}

\title{\plaintitle}

\numberofauthors{1}
\author{%
  \alignauthor{Haojian Jin, Vita Chen, Ritwik Rajendra, Jason Hong\\
    \affaddr{Carnegie Mellon University, Pittsburgh, PA}\\
    \email{haojian@, jasonh@cs.cmu.edu, vitac215@, ritwik.rajendra8119@gmail.com}}\\
} 

\maketitle

\input{parts/p0-abstract}

\input{parts/p1-intro}

\input{parts/p2-related_work}
\input{parts/p3-implementation.tex}

\input{parts/p4-applications}

\input{parts/p5-discussion}
\bibliographystyle{acm-sigchi}
\bibliography{proceedings}

\end{document}

%% file: parts/p0-abstract.tex
\begin{abstract}

Email is one of the most successful computer applications yet devised. 
Communication features in email, however, have remained relatively static in years.
We investigate one way of expanding email functionality without modifying the existing email infrastructure. 
We introduce \textit{email late bound content}, a simple and \textit{generalizable} technique that defers message content binding through image lazy-loading.
Parts of an email are converted into external images embedded in HTML code snippets, making it so that  email clients will defer the image download (i.e. content binding) until the moment users open the email. 
This late bound content allows email senders and third party services to update delivered emails. 
To illustrate the utilities of late bound content, we present four new example features and discuss the tradeoffs of email content late binding.

\end{abstract}



%% file: parts/p1-intro.tex
\section{Introduction}


Email is an integral form of communication, connecting an estimated 3.7 billion people by 2017~\cite{emailstatsreport}. 
The great success of email can be related to a number of unique characteristics such as being asynchronous~\cite{thomas2006reconceptualizing}, 
textual~\cite{tyler2003can}, and efficient~\cite{renaud2006you}.
However, there have been few new communication features in email in past years, in contrast to new forms of messaging
that have been invented to accommodate users' evolving communication needs, such as self-destruct (e.g. SnapChat)
or post editing after sending (e.g. Slack). 


A major challenge to expand email functionality is to incorporate new features while maintaining the 
backward compatibility with the dated email delivery protocol design.
Today's email systems are based on a "store-and-forward" model~\cite{resnick2008internet}. 
When a user composes a message, the sender interface will upload the message to an SMTP (Simple Mail Transfer Protocol) server. 
The SMTP server will then copy the message to the recipient's email server. 
One implication of this design is that users lose the ability to directly update an email after it is sent out. 

Researchers have proposed several potential solutions to extend email. 
For example, in 2009, Huawei Technologies proposed an SMTP Service Extension for Message Recall to the Internet Engineering Task Force~\cite{SMTPServ47:online}.
Many Chrome Extensions~\cite{garfinkel2005make,googleen6:online, hanson2003method, Ruoti:2016:PWS:2984511.2984580, Snapmail2:online} have also been developed to support additional email delivery protocols. 
However, it is difficult to make changes like these, 
in large part due to the strong need for backward compatibility with the large installed base of existing email servers. 
Our approach explores a different point in the design space through \textit{dynamic images in HTML-based emails}, offering less functionality but is also simpler and more compatible with today’s large existing base. 

\begin{figure}
	\includegraphics[width=1.0\columnwidth]{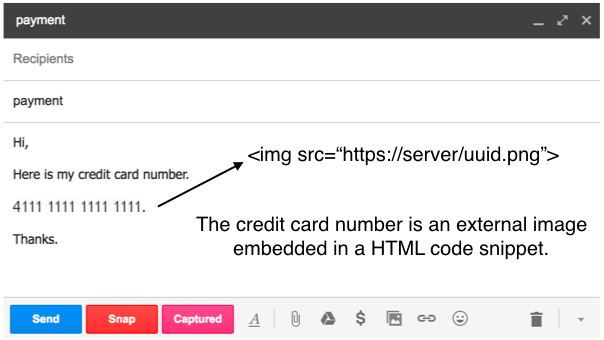}
	\caption{The concept of content late binding. 
		The presentation of Late Bound Content can be integrated into any emails. Email recipients are not required to install any software to view the content.}
	\label{fig:one}
\end{figure}


In the past, image display behaviors vary significantly across platforms. 
For example, some platform blocks images by default and users need to click "display images below" button to load.
Dynamic images are only used to display inessential information, such as a countdown timer~\cite{Dynamict54:online,Building46:online}.
Applications using dynamic images are extremely limited. 
Recently, we observe a convergence of image display behaviors (Table.~\ref{directdisplay}), which allows us to rethink the dynamic image applications in a more generalizable perspective.

In this paper, we abstract the dynamic images into a generic concept "late bound content" and illustrate the new possibilities through four new example functionalities. 
Our core idea is to use
references to images as \textit{late bound content}. Figure~\ref{fig:one} shows one example of content late binding. User-selected parts of an email can be converted into images that are referenced in an email, in this case a
credit card number. The sender can then update the late bound content in delivered emails by modifying the external images, or even put time limits so that the content can no longer be viewed after a certain period.
Our approach is backward-compatible and only requires changes on the sender's side.

We developed a Google Chrome Extension to help users create and manage the late bound content, which currently
works with Gmail and Outlook (previously known as Hotmail). We tested our approach on multiple platforms and found a high degree of compatibility, 
meaning that users can view this late bound content in the same way as regular emails. We also explore four different features for email, including 
1) self-destructing content, 2) continuous editing, 3) information dashboards, 4) real-time web references. We close with a discussion of tradeoffs in our proposed approach.

%% file: parts/p2-related_work.tex
\vspace{-0.1in}

\section{RELATED WORK}

We have organized past work into three categories: server and client based, clients-only based, sender client only based. 

One common approach is modifying both email servers and email clients. 
One example is email recall feature in Microsoft Outlook and Microsoft Exchange. 
Microsoft Outlook lets senders retract messages before recipients have seen the email. 
However, senders and receivers need to be on the same Exchange server and use Microsoft Outlook as their email clients~\cite{WhyYouCa99:online}. 
More recently, Gmail is incorporating~\cite{Bringing28:online}
the Accelerated Mobile Pages to make interactive and actionable email experiences. 
The major drawback to this approach is lack of compatibility with the large installed base of email servers and clients.

An alternative approach is modifying email clients through extensions. 
Zaplet~\cite{hanson2003method} is a widely deployed commercial email extension back in 2000. 
Zaplet embeds small Java applets in the email body and uses iFrame/iLayer/chart image elements to establish client-server bidirectional communication. 
Users can interact with the email like a micro webpage (type text response, click button, etc.). 
The embedded Java applets retrieve latest reference URLs and update the email content. More recently, 
numerous browser extensions have been developed to offer end-to-end email encryption ~\cite{googleen6:online, garfinkel2005make, Ruoti:2016:PWS:2984511.2984580}. As another example, SnapMail~\cite{Snapmail2:online} offers self-destructing emails by hosting
messages on external web pages and sending emails that only contain reference URLs.
These solutions require both senders and receivers to have the same email clients or extensions, which may negatively impact compatibility.

A complementary approach is to add additional features for senders that also aims for backwards compatibility. 
One example is the HTML-based email tracking, where the HTML specifies images that are loaded dynamically ~\cite{bouguettaya2003privacy}. 
Each time someone opens an email, the email client will request the image file from a web server,  creating a logged event that can capture who viewed an email and how many times it was viewed. 
There is value for end users and digital marketers in installing various plugins and services to enable tracking~\cite{EmailTra88:online, HubSpotI51:online}. 
Our approach falls into this category since we only require senders to install our extension.

%


%% file: parts/p3-implementation.tex
\section{Implementation}~\label{sec:imple}

\begin{table}
        \centering
        \begin{tabular}{ccccc}
                & Gmail & Outlook & Apple Mail & Thunderbird\\
        Web     & $\vartriangle$  &   \checkmark & N/A & N/A  \\
        Desktop &   N/A   &  $\square$ & \checkmark     & \checkmark        \\
        Mobile App  & \checkmark   & \checkmark     & \checkmark     &  N/A  
        \end{tabular}
        \caption{Results of our evaluation of late bound content with major email clients. 
        Most clients correctly display the latest version of images inline when the user opens the email (\checkmark). 
        Due to aggressive caching, Gmail ($\vartriangle$) requires a web page refresh to replace previously loaded late bound content. 
        Outlook (Desktop) ($\square$) pulls the latest contents after unblocking remote images.
        Note: Outlook (Web) is previously known as Hotmail.}
        \label{directdisplay}
\end{table}

Our implementation has two major system components: a Chrome Extension that helps senders create and update late bound content, 
and a RESTful server that generates and manages the external images. 
Depending on the content binding context, our Chrome Extension allows various ways to select the original content and then replace it with late bound content. 
We offer more details in the next section.

Once a user specifies the content to be made late bound, our extension will inspect the corresponding visual parameters 
(such as raw text, font, viewport sizes, image URLs) and send these parameters to our server. 
The server will then render the text into images and return unique URLs for generated images. 
Finally, our extension will replace the selected content with late bound content through HTML code snippets.

Once bound, the extension will also handle senders' future interactions and request server actions on external images. The server may delete, modify, replace corresponding images.

A key challenge of our content late binding techniques is compatibility across different email clients (web interfaces, desktop applications, and mobile apps). 
In particular, (a) do clients display inline images by default?, and (b) are there any effects with caching that may affect what a user sees when viewing an email for the first time or subsequent times?

Email clients have three primary ways to handle embedded images: inline display, fully blocked, and downloadable attachment. 
Most clients determine the strategy based on two image properties: 
1) if the image is hosted on a trustable (HTTPS) website; 
2) if the image file size and resolution are below certain thresholds.


Email clients also cache images differently. 
We found that most web interfaces do not cache embedded images and always download images from remote links directly.
The exception is Gmail, which first downloads external images to a Google server and then changes the "src" property of embedded images to the cached addresses.


We empirically tested different file sizes (from 1KB to 1MB) and resolutions (from 1 x 1 to 1000 x 1000) with various email clients.
We finally opted to limit the file size and resolution of each generated image to be under 200 kb and 299 x 524 respectively, 
which are the safe bounds for all our tested clients (Table~\ref{directdisplay}).  
If the content is too large to fit these constraints, we render the text as multiple images. 
We also hosted all images on an HTTPs website and set the cache duration to 0 in the HTTP response header.

We tested the compatibility of late bound content across major email clients with default configurations (Table~\ref{directdisplay}). 
The only exception is Outlook (Desktop), which blocks images by default and show a "download pictures" button. 
The other clients always display the latest late bound contents inline when the user opens the email.

%% file: parts/p4-applications.tex
\section{Example Extensions}
  To help motivate our idea and illustrate the potential of late bound content, 
  we devised four types of \textit{proof-of-concept} example extensions.
  All the use cases have been tested in the clients described above.

\begin{figure}
  \includegraphics[width=1.0\columnwidth]{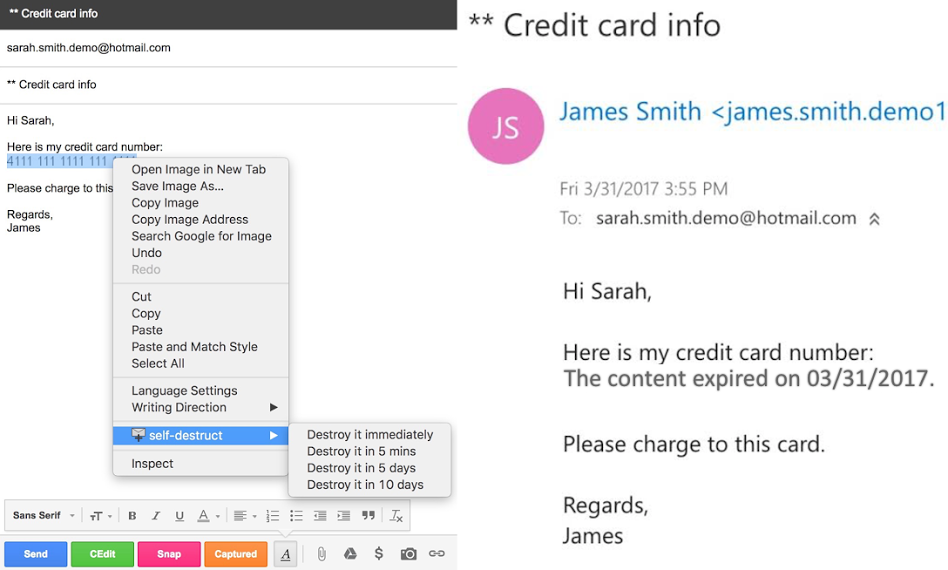}
  \caption{Example of Self-destructing email. Left: A Gmail user writes an email containing sensitive information and specifies the condition to expire. Right: An expiration notification has replaced the confidential information in the recipient's email (Hotmail).}
  \label{fig:self-destruct}
\end{figure}


\subsection{Self-destructing Content in Emails}
One use of late bound content is to send sensitive information that has a limited time to live.
For example, a user may need to send her credit card number to a trusted merchant for payment. 
However, the merchant might not delete the email afterward. If an attacker gains access to the merchant's
email account, the user's credit card number would be exposed.

Our Chrome extension provides two ways to create self-destructing content. 
Users can either select text manually or check on the auto extraction option.
In auto extraction mode, we use a text scrubbing library~\cite{scrubadu35:online} to automatically identify and convert sensitive information into late bound content before sending.

Users can also specify parameters for the self-destruct behavior, in terms of an expiration date, the number of times the content can be viewed, or a combination (Fig.~\ref{fig:self-destruct} left). 
For example, the users can ask the server to destroy the late bound content three days after the first time the content is viewed.
Our server counts the view times by tracking the number of images downloading. Once the self-destruct condition is satisfied, the server will replace the original image with a self-destruct notification message (Fig.~\ref{fig:self-destruct} right).

Another important design space of late bound content is the extension of kinetic typography (KT). KT refers to the art and technique of expression with animated text~\cite{lee2002kinetic}. Most successful KT applications~\cite{ford1997kinetic, ishizaki1998kinetic} can be applied to email text directly by converting text to animated images (i.e. GIFs), e.g., expressing emotions, directing reader attentions. For example, small vibrations on the keywords can be used to convey affective content with high levels of arousal, such as excitement or anger.

\begin{figure}
	\includegraphics[width=1.0\columnwidth]{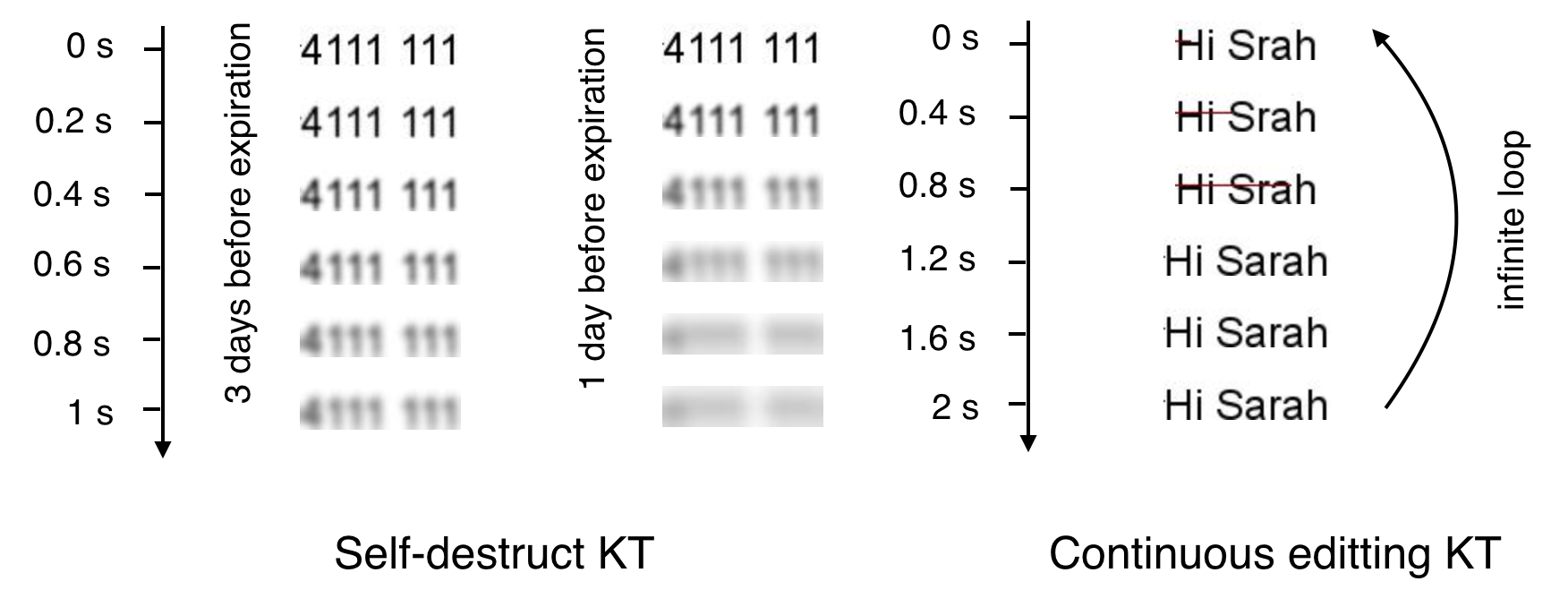}
	\caption{Animated image visualizations of Kinetic Typography. The static images are mapped to the corresponding timestamp in the animation. Self-destruct KT examples (left) illustrate two examples at 3 days and 1 day before expiration, respectively). 
		The continuous editing KT example (right) illustrates the modification history described in Fig.~\ref{fig:post-editing}. All animated images loop repeatedly.}
	\label{fig:kinetic-typo}
\end{figure}

Applying KT on late bound content further expand the email feature space. 
The self-destruct KT extension (Fig.~\ref{fig:kinetic-typo} left) can help differentiate the late bound content from the regular text and simulate the text aging process. This KT extension turns any self-destruct content into a blurring animation. As the time moves towards the expiration date, the final frame of the animation becomes increasingly blurred. We empirically set the animation length to one second for all the images and generate the animations at 10 frame-per-second. Our server updates the corresponding animated image every three hours.


\begin{figure}
	\includegraphics[width=1.0\columnwidth]{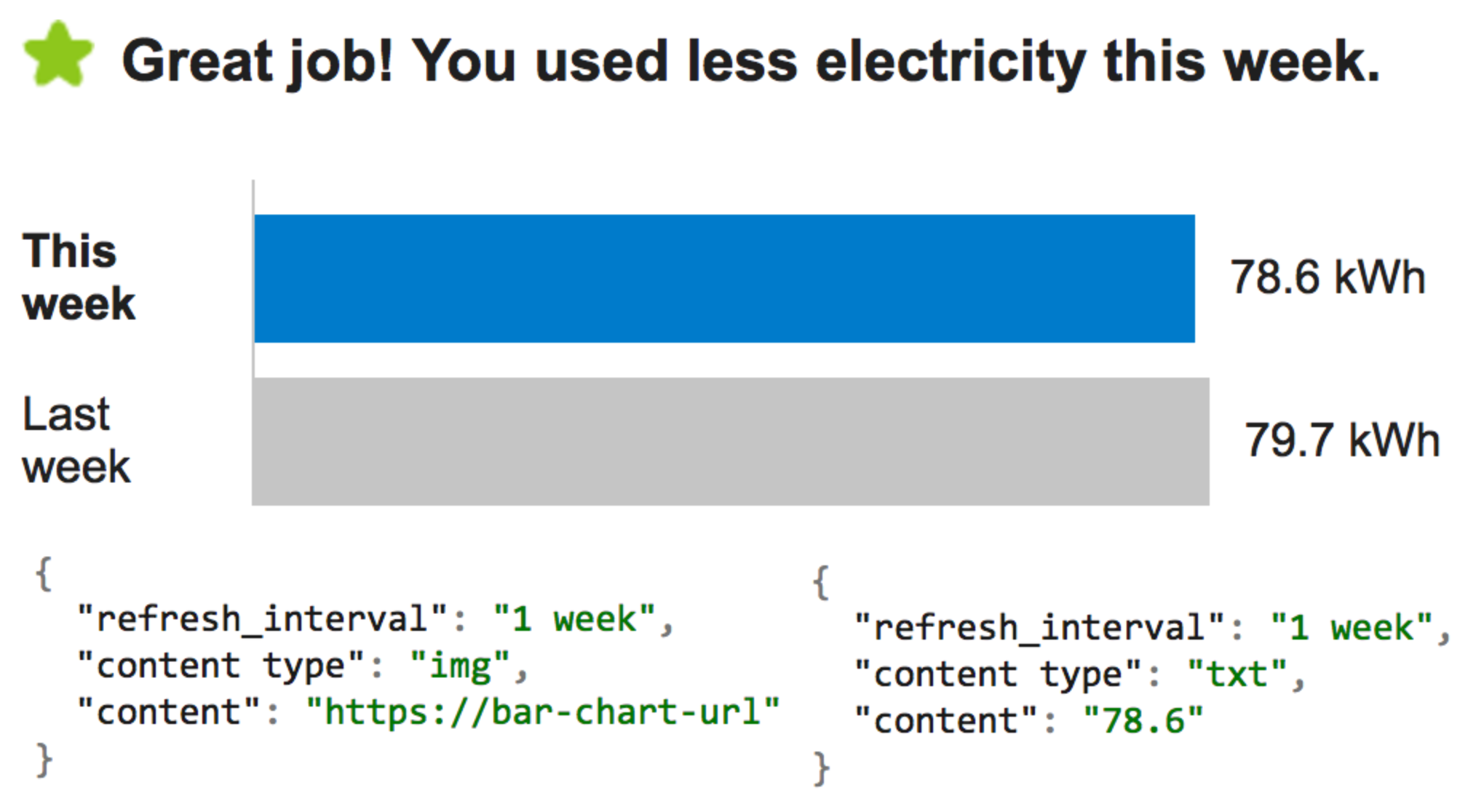}
	\caption{Reusable weekly electricity report. The chart and the numbers are bound to a corresponding RESTful services. The RESTful service specifies the refresh rate and content updates.}
	\label{fig:reusable_emails}
\end{figure}



\subsection{Information Dashboards}
Email overload is another common problem with email~\cite{szostek2011dealing}. 
Email users receive an average of 92 emails per day~\cite{emailstatsreport}. 
However, current estimates are that more than 60\% of the overall email traffic are machine generated (subscriptions, digital marketing, etc.)~\cite{ailon2013threading}. 

We propose a new approach, information dashboards, to reduce the number of similar subscription emails that are sent to users on a regular basis, which are often redundant since they share the same template~\cite{ailon2013threading}.
For example, two adjacent weekly electricity report emails share few differences except for the date, the power consumption numbers and the visualization based on the numbers. 
Converting these few changing contents into late bound content allows senders to send the subscription email only once and update the content subsequently.

Our Chrome Extension provides an easy way to create information dashboards: senders can copy-paste the original HTML email into the composing window and then bind images or texts to corresponding RESTful services.
Figure~\ref{fig:reusable_emails} shows an example of an information dashboard for weekly electricity report and two RESTful services.
Our server will contact the services and update the images at specified time intervals.
Without worrying about spamming, senders can now send the dashboard email once and show \textit{real-time} data (e.g. power consumption, Amazon Web Services real-time bill) in one email.

\setlength{\belowcaptionskip}{-10pt}

\subsection{Real-time Web References}
\begin{figure}
  \includegraphics[width=1.0\columnwidth]{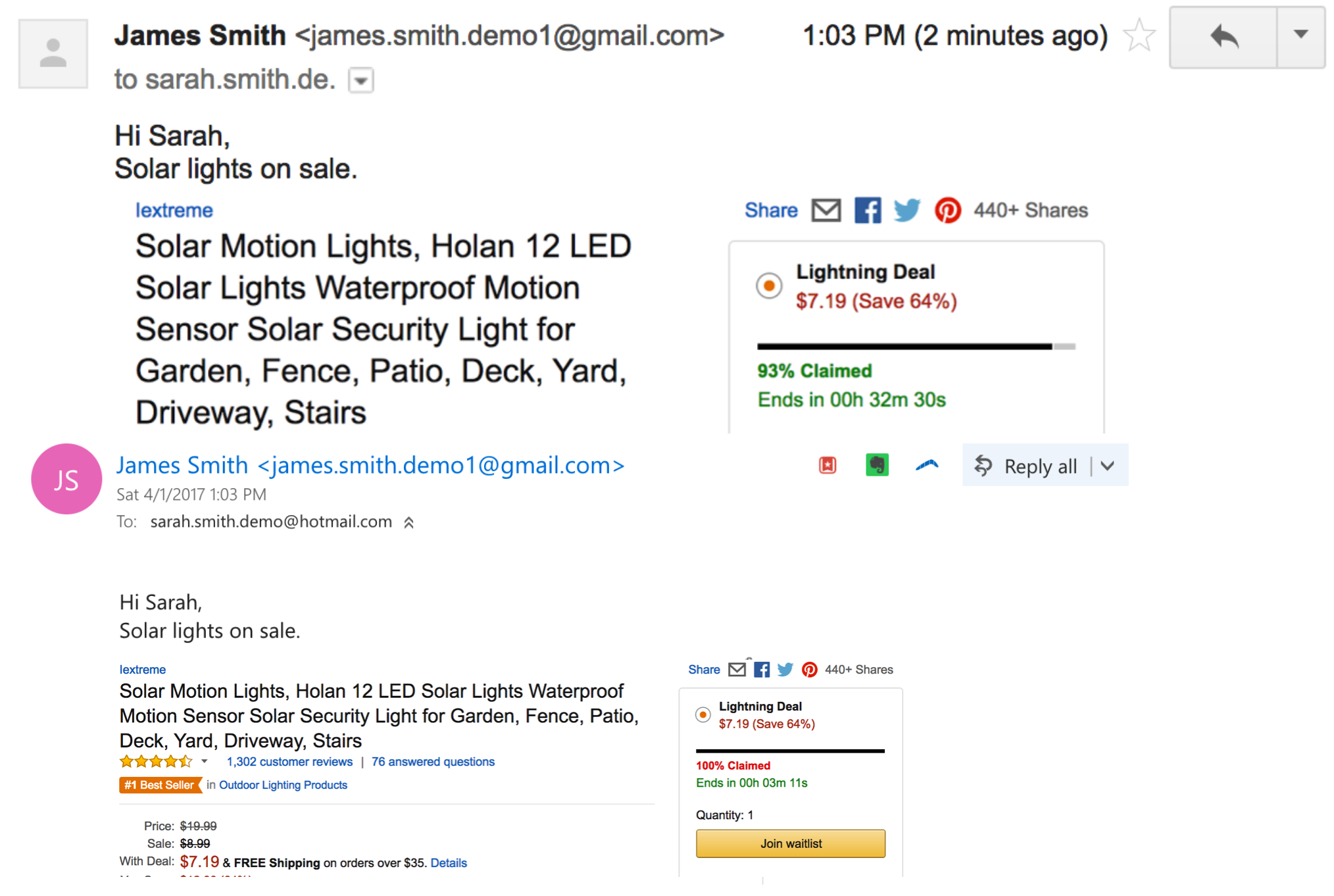}
  \caption{Example of Real-time Web References. A Gmail user shares an Amazon lightning deal screenshot with a Hotmail user. However, the deal has expired at the time the recipient opens the email (i.e. 100\% claimed).}
  \label{fig:realtimewebreference}
\end{figure}

There are many situations when users send out web screenshots in emails~\cite{cecchinato2016finding}. For example, a user might email a web screenshot to share a popular product sale with her friend  (Fig.~\ref{fig:realtimewebreference}). However, the product may already be out of stock by the time the recipient opens the email. 

Late bound content can be used to send close to real-time screenshots. 
Our Chrome extension allows users to screenshot parts of a web page and then insert it into the email body as late bound content. 
The server will then fetch the latest screenshots and replace the corresponding images periodically. 
When a user opens the email, the email should contain the most recent screenshot. Using the sale example above (Fig.~\ref{fig:realtimewebreference}), the screenshot would show whether the product is still available or out of stock.


Our current implementation can only track the public websites which do not require login or user interactions. 
By integrating an authentication system, we can potentially track password-protected websites as well.

\subsection{Continuous Editing}
Typos~\cite{boland2016if} and unintended messages~\cite{carvalho2007preventing} are common problems. 
Newer messaging applications offer features for addressing these issues. 
For example, Slack lets senders retract and edit sent messages. WeChat lets senders retract messages in the first 2 minutes after sending. 

\begin{figure}
  \includegraphics[width=1.0\columnwidth]{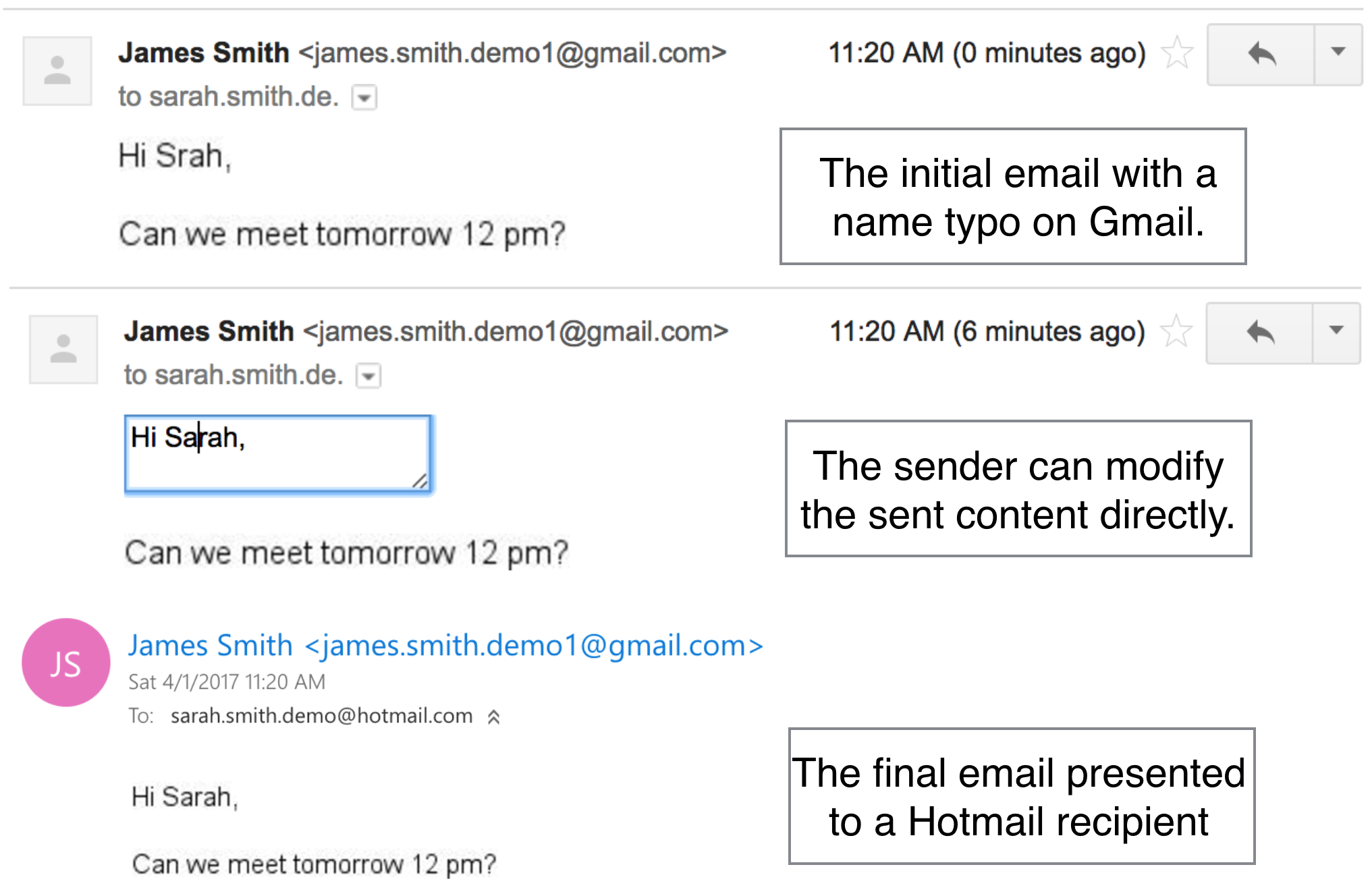}
  \caption{Example of Continuous Editing. A Gmail user writes an email to a Hotmail user with a typo in the recipient's name.}
  \label{fig:post-editing}
\end{figure}

Late bound content offers the ability to continuously edit regular emails after being sent (e.g., short mobile email replies and self-emails). 
Senders can click late bound content to enter the editing mode and modify the content directly (Figure~\ref{fig:post-editing}). 
Once they finish editing, the server will update the corresponding images immediately.

We currently only allow senders to modify sent messages until the recipient opens the email. 
Our extension generates a unique hash key for each piece of late bound content when created.  Then the extension saves this key in a browser cookie and on the remote server database. 
We use this key to differentiate senders from recipients and to determine if the current user has an editing access. 
The server will expire the hash key once the recipient opens the email.

We also develop an KT extension to alleviate users' concerns about the archival nature conflict.
The KT extension records the modification history and will replay the change history in an animated image. 
As illustrated in Figure~\ref{fig:kinetic-typo} (right), the animated image first plays strikethrough animations to the obsolete text (0-1s) and then shows the most recent version (1-2s). 

%% file: parts/p5-discussion.tex

\section{Discussion}

Although email is a decades-old application, new trends and technologies allow us to think about email in exciting new ways that were not possible in the early days. 
Our implementation of late bound content is feasible mainly because major email clients have adopted techniques including HTML-based email, image lazy-loading, and browser cookie access. 
Here we further present a thorough discussion of various potential tradeoffs of late bound content.



\textbf{Example use cases}.
In general, late binding is more useful for short-lived content than long-lived content, such as coupons, deals, and news/local updates. Recent study~\cite{bentley2017if} shows that “receiving advertisements/coupons/deals from stores” is the \#1 use of email (67\%),  "news/sports updates" (15\%) and "local updates" (11\%) are top categories as well. 
Cecchinato et al.’s qualitative diary study~\cite{cecchinato2016finding} finds that much email content was obsolete once used. 
Bao et al.~\cite{bota2017self} studied one type of short-lived email (self-emails) and found that the majority of self-sent emails have relatively short lifetimes (80\% less than a month).

One simple use of IDash can be a more accessible \textit{real-time} AWS billing dashboard.  If a user pins the AWS billing email to the top of Inbox (a feature in Hotmail), the user can quickly access the real-time billing information and notice abnormal usages. Similar usages can potentially enable a new email interaction paradigm that shifts the email consumption behavior from “push” to “pull.” 

Continuous editing (CE) also can be beneficial in many use cases, such as short mobile email replies and self-emails. Bao et al.~\cite{bao2011smart} found users often send short replies on mobile and make sure to include "sent from my iPhone" disclaimer to explain typos. CE offers an opportunity for fixing typos later on. CE can also apply to reusable self-emails, which are often used as reminders~\cite{bellotti2003taking} and to-do lists~\cite{bota2017self,mcdowell2004semantic}.

RWR also can turn homepages like Yahoo News into a partial screenshot. Everytime the user opens the email, she can see the latest news. Similar examples apply to the stock info, traffic status, etc.



\textbf{Dynamic emails management}.
Dynamic emails (e.g. IDash and RWR) require users to check dashboard consciously. As a result, users may need to develop new email management strategy. Past research found participants used a variety of workaround strategies to make certain email messages more conspicuous~\cite{cecchinato2016finding}. 
For instance, “Pin at Top” (a standard feature in many clients) can be applied to the most important dashboards. Users can organize other emails through folders/tags/stars. 
Bentley et al.~\cite{bentley2017if} found that their participants subscribed to an average of 93 different email lists. 
Applying IDash to some of the subscriptions could significantly reduce the number of emails, 
and the resulting number of dashboards should be manageable.

\textbf{Sending email by reference}.
An alternative approach to late bound content is hosting the email body on an external website and sending emails containing only the web links~\cite{Snapmail2:online}. 
This approach makes email a notification-only communication tool instead of a personal information management tool~\cite{ducheneaut2001mail}, which conflicts with common user habits. 
Moreover, frequently switching between external websites and email clients can make the email overload even worse.

\textbf{Edit permission management}.
We implement all applications using browser cookies. This design is the tradeoff for the deployment cold start. Users don't need to register and login, but this design also prevents cross machine editing. SD and CE senders can only modify the late bound content using the same machine on which the content was created. Incorporating an account system can address this limitation.

\textbf{Image cache}.
Table~\ref{directdisplay} shows that all the tested clients will only pull the latest version of images when the user opens the email. In fact, the content can be updated even if the user keeps the browser tab open infinitely without refreshing, since most browsers invalidate non-active tabs to save memory~\cite{TabDisca0:online}. 

\section{Limitations}

\textbf{Text to Image Conversion}.
Converting text into images also consumes extra bandwidth (SD/CE/IDash). 
We ran a test that renders 1000 100-character strings into images with default Gmail fonts, producing an average image size of 7.4 KB before compression.
Considering emails are short in length (< 60 words~\cite{kooti2015evolution}), the bandwidth difference is negligible with current LTE/WiFi network speed. 

Our experiments find some email clients modify the image display style. The image can be blurred if the size doesn't match exactly. A more sophisticated image hosting server can address that issue.

\textbf{Search and other raw text features}.
One major drawback of late bound content is that it will break functionalities that rely on raw text, such as email search, text copy-paste, accessibility, and automatic event extraction.
We considered putting backup text into "alt" tags, but this might lead to confusion if the updated late bound contents no longer match with the original "alt" text.

These limitations should be considered in the final application design. For example, IDash, RWR, and SD only turn key numbers/phrases into images, so this would impact some but not all search queries. CE should be applied to mainly short-lived content, which requires less searchability.

Better integration of optical character recognition techniques should help this problem in the future.
Another issue is that the display of late bound content depends on Internet accessibility. If a user is on a flight (and thus does not have access to the Internet), the late bound content might be out of date.

\textbf{User adoption study}.
We considered running an in-lab user acceptance test. However, our preliminary survey shows that users have developed various tactics and preferences to use emails. Running a small scale study may lead to bias~\cite{HTMLEmai65:online,WhyHTMLi65:online}. Instead, we focus on how our approach can be employed and articulate example applications to help motivate our ideas.

Besides, while these extensions are derived from the same concept technically, the final presentation and usages are quite different from the non-technical end-user perspectives. The final unified system may only contain parts of the extensions described above.

\section{Conclusion and Future Work}

Late binding is a computer programming mechanism in which the object binding is deferred until runtime. 
We apply this concept to the email content and present five types of novel use of the \textit{email late bound content}. These new applications can potentially improve the email's security, usability, and email content management. 

For future work, we plan to release this project on Chrome Extension Store and apply 
 late bound content to rich media. For example, we can bind GIFs to summarized security camera footage and the time-lapse of traffic visualization in Google Maps. 
We believe late bound content can be deployed publicly as a standard feature of commercial emails, providing new ways to interact with email.